\documentclass[11pt,a4paper]{article}
 
\usepackage{graphicx}
\usepackage{epstopdf}
\usepackage{jcappub}
\usepackage{amsmath,amsthm,latexsym,amssymb,amsfonts,epsfig}
\usepackage{fontenc,dsfont,layout}
\usepackage{hyperref}
\usepackage{psfrag}
\usepackage[cp1250]{inputenc}
\usepackage[usenames,dvipsnames]{xcolor}

\numberwithin{equation}{section}

\title{Multi-phase induced inflation in theories with non-minimal coupling to gravity}
\author[a]{Micha{\l}~Artymowski}
\author[b]{Zygmunt~Lalak}
\author[c]{Marek~Lewicki}

\affiliation[a]{Institute of Physics, Jagiellonian University\\
{\L}ojasiewicza 11, 30-348 Krak{\'o}w, Poland}
\affiliation[b,c]{Institute of Theoretical Physics, Faculty of Physics, University of Warsaw\\ 
ul. Ho\.{z}a 69, 00-681 Warszawa, Poland}

\emailAdd{Michal.Artymowski@uj.edu.pl}
\emailAdd{Zygmunt.Lalak@fuw.edu.pl}
\emailAdd{Marek.Lewicki@fuw.edu.pl}

\abstract{In this paper we investigate the induced inflation with two flat regions: one Starobinsky-like plateau in big field regime and one shorter plateau around the saddle point of the Einstein frame potential. This multi-phase inflationary scenario can be used to solve the problem of classical cosmology. The inflation at the saddle-point plateau is consistent with the data and can have arbitrarily low scale. The results can be useful in the context of the Higgs-Axion relaxation and in a certain limit they are equivalent to the $\alpha$-attractors.}


\begin{document}
\maketitle

\section{Introduction} \label{sec:Introduction}

Cosmic inflation \cite{Lyth:1998xn,Liddle:2000dt,Mazumdar:2010sa} is the well established theory of the early universe with good consistency with the data \cite{Ade:2015lrj}. It predicts the accelerated expansion of space together with flat power spectrum of primordial inhomogeneities of cosmic microwave background, which is measured by several experiments. Especially the latest data from the PLANCK experiment puts stronger constraints on the tensor-to-scalar ratio $r$, which tells us about the amplitude of primordial gravitational waves and about the scale of inflation. The PLANCK results favours the plateau-like potentials (for which the energy density of the potential is suppressed from above by the scale of inflation) over the big field models, like e.g. $m^2 \phi^2$ or other chaotic inflationary models (for which the potential is not limited from above). 
\\*

Examples of models with plateau are e.g. Starobinsky inflation \cite{Starobinsky:1980te,Barrow:1988xh}, Higgs inflation \cite{Bezrukov:2007ep}, and their generalisations \cite{Codello:2014sua,vandeBruck:2015xpa,Artymowski:2014gea,Ben-Dayan:2014isa,Motohashi:2014tra}. In addition, in Ref. \cite{Kallosh:2013hoa} the authors claim that in a particular form of supergravity the flat plateau appears for almost any scalar potential, which suggests that inflation from the plateau may be natural to obtain. In this paper we want to investigate one of the most general forms of a potential with a plateau, namely the induced inflation \cite{Kallosh:2013tua,Kallosh:2014laa,Giudice:2014toa,Kallosh:2014rha}. In this class of models the relation between non-minimal coupling to gravity $f(\varphi) R$ and a Jordan frame potential $U(\varphi)$ provides the flat region of the Einstein frame potential in the $f \gg 1$ limit. 
\\*

In Ref. \cite{Ijjas:2013vea} the authors argue that the suppression of plateau-like inflationary potentials leads to serious fine-tuning of initial conditions for the pre-inflationary universe. The reason for that is the following one: if one assumes the Planck scale as the scale of initial conditions of the universe, then the Einstein frame potential cannot have any significant contribution to initial energy density. Simply, the scale of the plateau is at least 10 orders of magnitude smaller than $M_p^4$. In order to avoid the domination of inhomogeneities, the energy density of which decreases slower than the time derivative of the field or the radiation energy density, one needs to assume that in the pre-inflationary universe there was a homogeneous region which consisted of  $\sim10^9$ causally disconnected Hubble horizons. 
\\*

A possible solution to that issue could be the additional inflationary era around the Planck scale, which would smooth out the Universe and therefore prepare it for inflation around the GUT scale. This approach can be seen in Ref. \cite{Hamada:2014wna}, where the pre-inflationary phase is generated by topological defects. Another example of a potential, which could solve the problem of initial conditions for inflation comes from $\alpha$-attractors \cite{Carrasco:2015rva}, where besides the Starobinsky plateau one obtains the second plateau, possibly at the Planck scale. In both cases the first phase of inflation happens close to the Planck scale, so there is no hierarchy gap between scales of inflation and initial conditions. In this paper we obtain a similar result using induced inflation with two plateaus.
\\*

The arguments mentioned above are not fully accepted by scientific community. For instance in Ref. \cite{Gorbunov:2014ewa} the Authors prove that initial conditions set in the Planck scale of the Jordan frame do not lead to the energy gap and fine tuning of initial conditions. Also in \cite{Carrasco:2015rva,Linde:2004nz,Linde:2014nna} one finds several arguments against statements listed in Ref. \cite{Ijjas:2013vea}. Therefore we want to stress that besides solving the issue of initial conditions for inflation our primary motivation is seeking for generalised forms of previously analysed examples of induced inflation.
\\*

In what follows we use the convention $8\pi G = M_{p}^{-2} = 1$, where $M_{p} = 2.435\times 10^{18}GeV$ is the reduced Planck mass.
\\*

The structure of this paper is as follows. In Sec. \ref{sec:general} we introduce the issue of induced inflation and the possibility of obtaining a saddle point inflation in this class of scalar-tensor theories. We also discuss the equivalence of this model to $\alpha$-attractors with non-minimal coupling to gravity. In Sec. \ref{sec:BD} and \ref{sec:Higgs} we investigate the Brans-Dicke-like and Higgs inflation-like induced inflation with the saddle point. In Sec. Finally we summarise in Sec. \ref{sec:Summary}.

\section{Saddle point inflation for general form of non-minimal coupling} \label{sec:general}

\subsection{Conditions for the existence of the saddle point}

Let us assume that the Universe is described by the flat FRW metric tensor with the following Jordan frame action of a scalar-tensor theory
\begin{equation}
S = \int d^4 \sqrt{-g} \left[\frac{1}{2}f(\varphi)R + \frac{1}{2}(\partial\varphi)^2-U(\varphi)\right] \, ,
\end{equation}
where $U(\varphi)$ is the Jordan frame scalar potential. In \cite{Kallosh:2013tua,Kallosh:2014laa,Giudice:2014toa,Kallosh:2014rha} it was shown that for a general $f(\varphi)$ one should obtain a flat plateau of the Einstein frame potential for
\begin{equation}
U = M^2 (f-1)^2 \, .
\end{equation}
Due to the non-minimal coupling to gravity it is convenient to transform the action into the Einstein frame, where the gravitational part of the action obtains its canonical form. Let us define the Einstein frame metric tensor $\tilde{g}$ and scalar field $\phi$ by
\begin{equation}
\tilde{g}_{\mu\nu} = f(\varphi) g_{\mu\nu} \, , \qquad \frac{d\phi}{d\varphi} = \sqrt{\frac{3}{2}\left(\frac{f_\varphi}{f}\right)^2+\frac{1}{f}} \, ,
\end{equation}
where $f_\varphi := \frac{df}{d\varphi}$. The Einstein frame action reads
\begin{equation}
S_{EF} = \int d^4 \sqrt{-\tilde{g}} \left[\frac{1}{2}\tilde{R} + \frac{1}{2}(\partial\phi)^2-V(\phi)\right] \, ,
\end{equation}
where $V$ is the Einstein frame potential
\begin{equation}
V = M^2 \frac{(f-1)^2}{f^2} = M^2 \left(1-\frac{1}{f}\right)^2 \, .
\end{equation}
For $f \gg 1$ one finds $V \sim M^2 = const$ and therefore $M$ is an energy scale of the plateau. Let us assume that the Einstein frame potential has a stationary point at any $\phi = \phi_s$. The stationary point could be in particular a local maximum or a saddle point, which could generate inflation from higher order corrections to a scalar-tensor theory. It is easy to show that besides the GR minimum, at which $f(\varphi) = 1$ and $V(\varphi) = 0$ one obtains
\begin{equation}
\frac{d^mV}{d\phi^m} = 0\quad \Leftrightarrow \quad \frac{d^m f}{d\varphi^m} = 0
\end{equation}
Thus for any $n$, where $\varphi^n$ is a highest order term in $f(\varphi)$, one obtains maximally $n-1$ constraints from the stationary point condition. The most popular form of inflation from stationary point is the saddle-point inflation \cite{Allahverdi:2006we}, which may be used to significantly lower the scale of inflation. Unfortunately the saddle-point inflation with only two derivatives vanishing on $\varphi_s$ is inconsistent with the data due to too small value of $n_s$. On the other hand it was shown that the flat region around the $m$-order stationary point may be a viable source of inflation \cite{Hamada:2015wea}, which is partially our motivation to consider this kind of potentials. This fact was already applied in the Ref. \cite{Artymowski:2015pna,Artymowski:2016mlh}, where the saddle-point inflation was investigated in the context of $f(R)$ theory.
\\*

During inflation one can use the slow-roll approximation, for which $\ddot{\varphi} \ll 3H\dot{\varphi}$ and potential terms dominate over kinetic ones. In such a case one can calculate the number of e-foldings defined by
\begin{equation}
N = \int H dt \simeq \int \frac{f-1}{2f_\varphi f}\left(f+\frac{3}{2}f_\varphi^2\right)d\varphi \, .
\end{equation}
In the $f_\varphi ^2 \gg f$ limit, which is the strong coupling limit at the Starobinsky-like plateau, one obtains $N \simeq 3(f - \log f)/4$. The slow-roll approximation is valid as long as 
\begin{equation}
\epsilon:= \frac{1}{2}\left(\frac{V_\phi}{V}\right)^2 \ll1 \, , \qquad |\eta|:= \left|\frac{V_{\phi\phi}}{V}\right| \ll 1
\end{equation}
From slow-roll parameters one constructs two observables, which can be compared with the data, namely the tensor-to-scalar ratio $r:=16\epsilon$ and the spectral index $n_s:= 1-6\epsilon+2\eta$. Note that for any $f$, which satisfies $f\gg 1$ and $f_\varphi ^2 \gg f$ one finds $N\simeq 3(f - \log f)/4 \simeq 3f/4$, $\epsilon \simeq 3/(4N^2)$, $\eta \simeq -1/N$, and $n_s \simeq 1-2/N$. This is the attractor behaviour in the strong coupling limit described in \cite{Kallosh:2013tua}.
\\*

\begin{figure}[h]
\centering
\includegraphics[height=4.5cm]{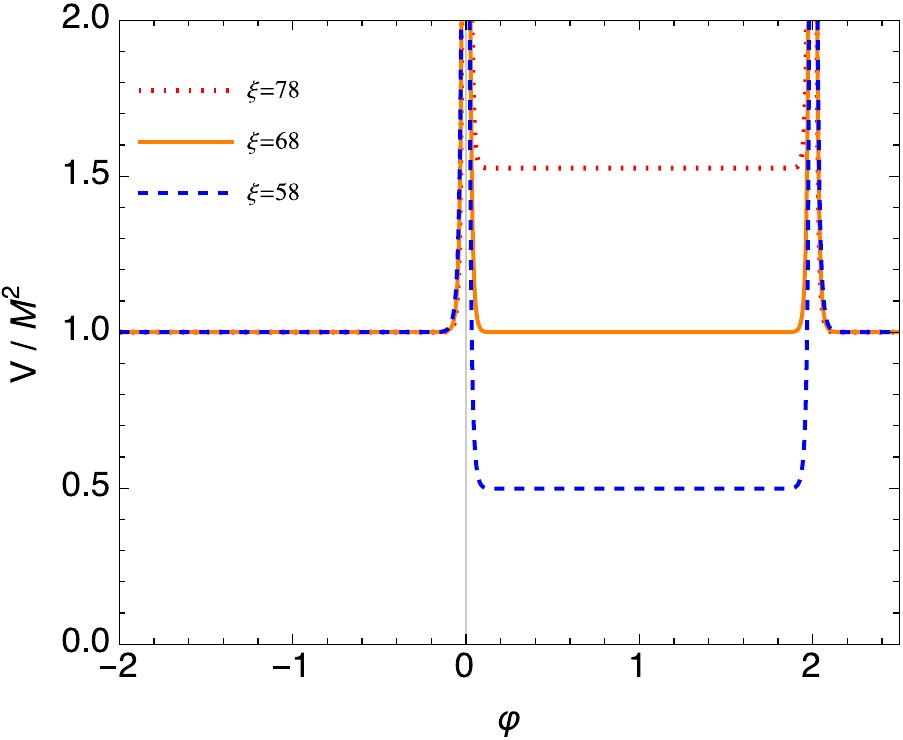}
\caption{\it The Einstein frame potential for even $n$. We have assumed $\lambda = \lambda_1$ and $\xi = 34$. The only way to obtain GR minimum is to assume that $\xi = n$, which still does not remove barriers of the potential separating the minimum from plateaus. There are several flat regions of the potential, but lack of the graceful exit makes it useless for inflation.} 
\label{fig:even}
\end{figure}

Let us consider the following form of $f(\varphi)$ 
\begin{equation}
f(\varphi) = \xi \sum_{k=0}^n \, \lambda_k \, \varphi^k \, ,\label{eq:lambdak}
\end{equation}
where $\lambda_k$ are constants. Note that $\varphi$ is expressed in Planck units $M_p = 1$ and therefore $f(\varphi)$ is dimensionless. One can always redefine the $\xi$ constant that $\lambda_1 = 1$. I such a case the demand of the existence of a $n-1$ order stationary point in the Einstein frame leads to the following form of $\varphi_s$ and $\lambda_k$ coefficients
\begin{equation}
\varphi_s =\left(n \, \lambda\right)^{\frac{-1}{n-1}}\, , \qquad \lambda_k = (-1)^{k+1}\frac{(n-1)!}{k!(n-k)!}(n \, \lambda)^{\frac{k-1}{n-1}} \, , \label{eq:lambdaksaddle}
\end{equation}
where $\lambda:= \lambda_n$ is a free parameter of the theory. The additional free parameter is $\Lambda:=\lambda_0$, which can take any real values, but we will restrict ourselves to $\Lambda = 0$ (Brans-Dicke-like case) and $\Lambda = 1$ (Higgs inflation-like case). We have now 4 independent parameters ($M$, $\xi$, $\lambda$, $n$) and only one of them can be removed by the normalisation of inhomogeneities. For simplicity let us choose some particular form of $\lambda$. The simplest choice of $\lambda$ would be $\lambda = 1/n$, since the $n \, \lambda$ coefficient is then equal to one. Nevertheless one can choose a different $n$-dependence of $\lambda$, which as we will show will lead to different forms of the Einstein frame scalar potential. Using Eq. (\ref{eq:lambdak},\ref{eq:lambdaksaddle}) one finds
\begin{equation}
f(\varphi) = \Lambda + \frac{\xi}{n}\left(n \, \lambda\right)^{\frac{-1}{n-1}}\left(1 + \left(\left(n \, \lambda\right)^{\frac{1}{n-1}}\varphi - 1 \right)^n\right) \, .\label{eq:fgeneral}
\end{equation}
For even values of $n$  there are 3 areas of the Einstein frame potential, which are flat: i) The stationary point, which is an additional vacuum with very flat potential around the local minimum. This minimum has non-zero vacuum energy density and it is separated from other regions by very step walls of the potential. ii) The Starobinsky-like plateau. iii) the flat area in the $f < 0$ regime, where the gravity is repulsive. Thus, for even $n$ there is no inflationary scenario, which allows the graceful exit and good low-energy limit of the theory. The even $n$ case for $\Lambda = 0$ is shown in the Fig. \ref{fig:even}
\\*

In what follows we shall restrict our analysis to odd $n$. In this case the repulsive gravity appears for $\varphi < 0$. Luckily the inflationary part of the potential as well as the GR vacuum are separated from negative $\varphi$ with the step slope of the Einstein frame potential at the $\varphi \to 0$ limit (see plots \ref{fig:Saddle1} and \ref{fig:Saddle2} for details). In $\varphi \gg 1$ limit (the Starobinsky plateau) one finds $f \simeq \xi\, \lambda \, \varphi^n$, so far on the plateau this model recovers results of \cite{Kallosh:2013tua}. The GR vacuum of the model is obtained for 
\begin{equation}
\varphi_{\text{\tiny GR}} = (n \, \lambda)^{\frac{-1}{n-1}}\left(1+ \left( \frac{n}{\xi}(n \, \lambda)^{\frac{1}{n-1}}(1-\Lambda) - 1 \right)^{\frac{1}{n}}\right) \, .
\end{equation}
For $\lambda =\lambda_1: =1/n$ and $\lambda = \lambda_2 : = \frac{1}{\xi}(\xi/n)^{n}$ the Eq. (\ref{eq:fgeneral}) simplifies into
\begin{eqnarray}
f(\varphi) = \Lambda + \frac{\xi}{n}\left(1 + \left(\varphi - 1 \right)^n\right) \, ,\qquad \varphi_s = 1\qquad &\text{for}& \qquad \lambda = \lambda_1 \, , \\
f(\varphi) = \Lambda + \left(1+ \left( \frac{\xi}{n}\varphi - 1 \right)^n\right) \, ,\qquad \varphi_s = \frac{n}{\xi} \qquad &\text{for}& \qquad \lambda = \lambda_2 \, . \label{eq:fn^-n}
\end{eqnarray}
Note, that  for $\xi = n$, $\xi > n$ and $\xi<n$ one obtains $\lambda_1 = \lambda_2$, $\lambda < \lambda_2$ and $\lambda > \lambda_2$ respectively. {Unlike the $\lambda = \lambda_2$ scenario for $\lambda = \lambda_1$ one obtains $\lambda_k \gg1$ for $n \gg 1$. This may lead to non-perturbative potentials and therefore $\lambda = \lambda_2$ seems to be preferred over $\lambda = \lambda_1$ case.} The Eq. (\ref{eq:fn^-n}) has a $n \to \infty$ limit, which is 
\begin{equation}
f = 1 + \Lambda -e^{-\xi \varphi} \, . \label{eq:ninf}
\end{equation}
This result holds even after multiplying $\lambda$ by any positive, n-independent constant. In this model the Einstein frame potential has an infinite wall at $\varphi = -\frac{1}{\xi}\log(1+\Lambda)$, which separates the repulsive and attractive gravity regimes. The GR minimum appears at $\varphi = -\frac{1}{\xi}\log(\Lambda)$ and therefore its existence requires $\Lambda > 0$. Thus only the Higgs inflation-like case shall be investigated in the $n \to \infty$ regime. Despite of the value of $\Lambda$ the saddle point moves to $\varphi \to \infty$ and inflation happens on the RHS of the GR minimum. This result corresponds to the one obtained in Ref. \cite{Artymowski:2015pna,Artymowski:2016mlh}, in which the authors demanded the existence of the stationary point of the $f(R) = \sum \alpha_n R^n$ model. Taking the number of terms to infinity also gave the saddle point that moved to $R \to \infty$.
\\*

{We have restricted ourselves to the particular form of scalar potential, which in the $f \gg 1$ limit always gives inflationary plateau in the Einstein frame. Nevertheless it was shown in Ref. \cite{Artymowski:2016pjz} that even in the model minimally coupled to gravity one obtains potentials with flat region around $\varphi_s$ for any $U = U(f(\varphi))$ . Therefore any $U(f)$ can create the inflationary plateau as long as $V(f(\varphi_s)) > 0$. Otherwise the flat region can be a minimum of $V$ or may have negative energy density. }

{\subsection{Equivalence to $\alpha$ attractors}} \label{sec:alphaattractors}

{In \cite{Artymowski:2016pjz} the Authors show that using $f$ as an effective scalar leads to a non-canonical kinetic term. In particular, for $f(\varphi)$ of the form of (\ref{eq:fgeneral}) one finds
\begin{equation}
(\partial \varphi)^2 = \frac{1}{\xi ^2}\left(\frac{\xi }{f \, n (\lambda \,  n)^{\frac{1}{n-1}}-\xi }\right)^{\frac{2 (n-1)}{n}} (\partial f)^2 \, .
\end{equation}
The kinetic term of has a pole at $f_p = f(\varphi_s)$ and therefore the pole corresponds to the stationary point of $f(\varphi)$. As shown in Ref. \cite{Galante:2014ifa,Kallosh:2013daa,Karananas:2016kyt} poles of kinetic term correspond to stretching of the scalar potential around the pole and therefore such theories are the natural source of flat inflationry potentials. For $\lambda = \lambda_2$ and in the $n \to \infty$ limit one finds
\begin{equation}
(\partial\varphi)^2 = \frac{(\partial f)^2}{\xi^2(f-1)^2} \, ,
\end{equation}
which is {very similar to} the kinetic term for of the $\alpha$-attractors  \cite{Carrasco:2015rva}, where the kinetic term of a scalar field $\psi$ is equal to $ (\partial \psi)^2\left(1-\frac{\psi^2}{6\alpha^2}\right)^{-2}$. To obtain the exact form from the $\alpha$-attractors kinetic term one needs
\begin{equation}
\psi(f) = \frac{\sqrt{6 \alpha } \left((6 \alpha  (1-f) \xi )^{\sqrt{\frac{2}{3 \alpha }} \xi }-1\right)}{(6 \alpha  (1-f) \xi )^{\sqrt{\frac{2}{3 \alpha }} \xi }+1} \, .
\end{equation}
Therefore our model {in $\lambda = \lambda_2$, $n \to \infty$ limit} is fully equivalent to the $\alpha$-attractor non-minimally coupled to gravity. This analysis would be still valid for any potential $U(f)$. The multi-phase inflation obtained in \cite{Carrasco:2015rva} is, on the conceptual level, identical with one one obtained by us. One flat region is obtained by assuming the existence of the flat potential (in our case it is a flat potential of induced inflation, in case of Carisso et. all it is a Starobinsky scalar potential), the other flat region is obtained by a stationary point or a pole of kinetic term, which as have shown, are equivalent to each other.}
\\*

\section{Brans-Dicke-like induced inflation} \label{sec:BD}

\begin{figure}[h]
\centering
\includegraphics[height=4.3cm]{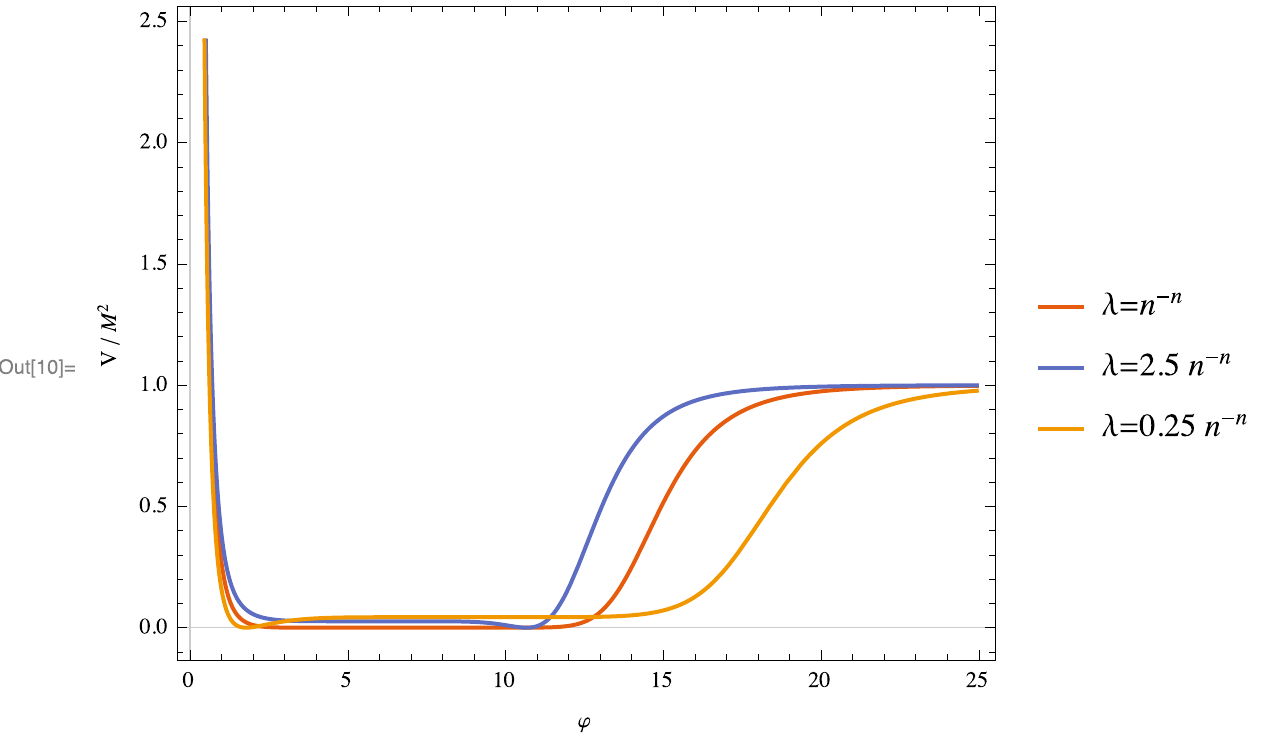}
\hspace{0.4cm}
\includegraphics[height=4.3cm]{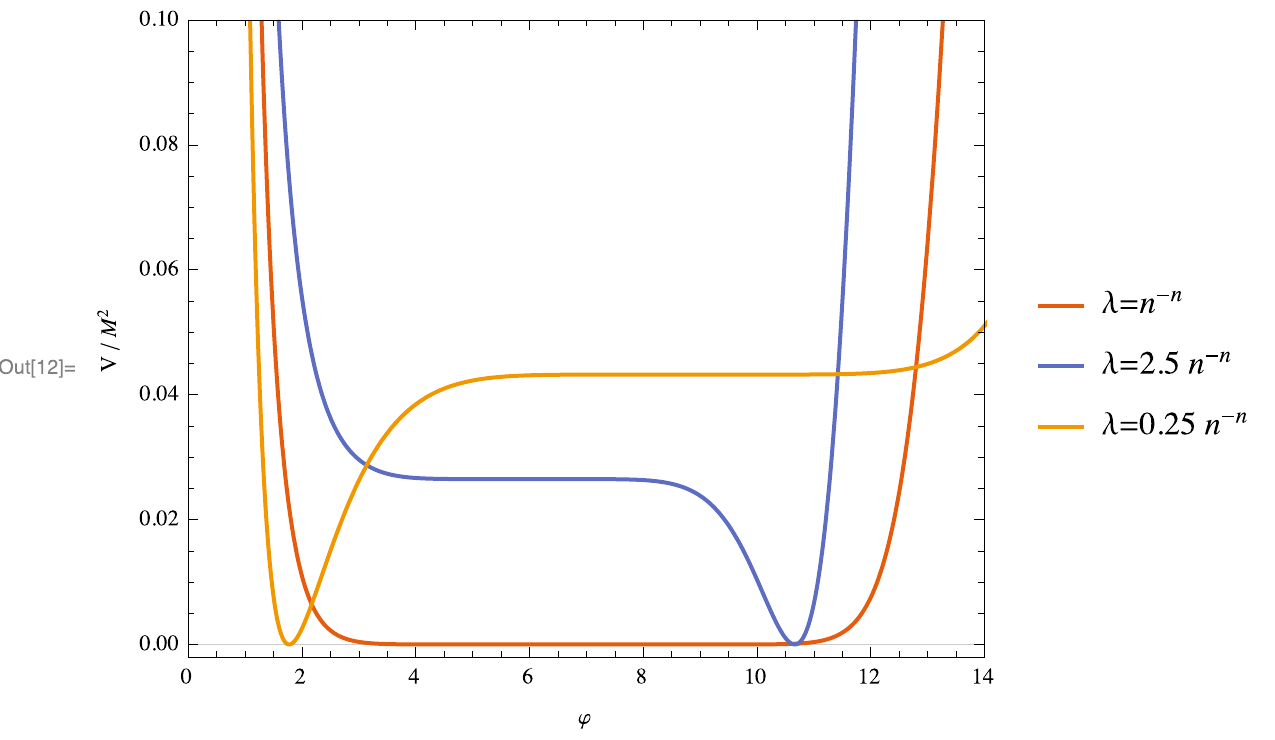}
\caption{\it Einstein Frame potential as a function of Jordan frame field for $\xi = 1$, $\lambda\sim \lambda_2$, $\Lambda = 0$ and $n=7$. The right panel zoom out the area of the GR minimum with possible additional saddle-point-like plateaus. The scale difference between  Note that $V $} 
\label{fig:Saddle2}
\end{figure}

In this section we consider $\Lambda = 0$, which makes the model alike the Brans-Dicke theory with higher order corrections. The Einstein frame potential has the following value at the stationary point
\begin{equation}
V(\varphi_s) = M^2\left(\frac{n}{\xi} (n \, \lambda)^{\frac{1}{n-1}}-1\right)^2 \, . \label{eq:VsBD}
\end{equation}
Thus, for every $\lambda > 2^{n-1} \lambda_2$ ($\lambda < 2^{n-1} \lambda_2$) one finds that the saddle-point plateau is higher (lower) than the Starobinsky-like one. For $\lambda = \lambda_2$ (or for $\lambda = \lambda_1$ and $\xi=n$) one finds $V(\varphi_s) = 0$, so the stationary point is a GR minimum. The main difference between this case and a regular induced inflation is a very wide and flat GR minimum, which decreases the mass of the scalaron in lower energies and makes it harder to screen modified gravity in lower scales. The $\lambda = \lambda_2$ case is presented in the Fig. \ref{fig:Saddle2}.
\\*

\begin{figure}[h]
\centering
\includegraphics[height=4.5cm]{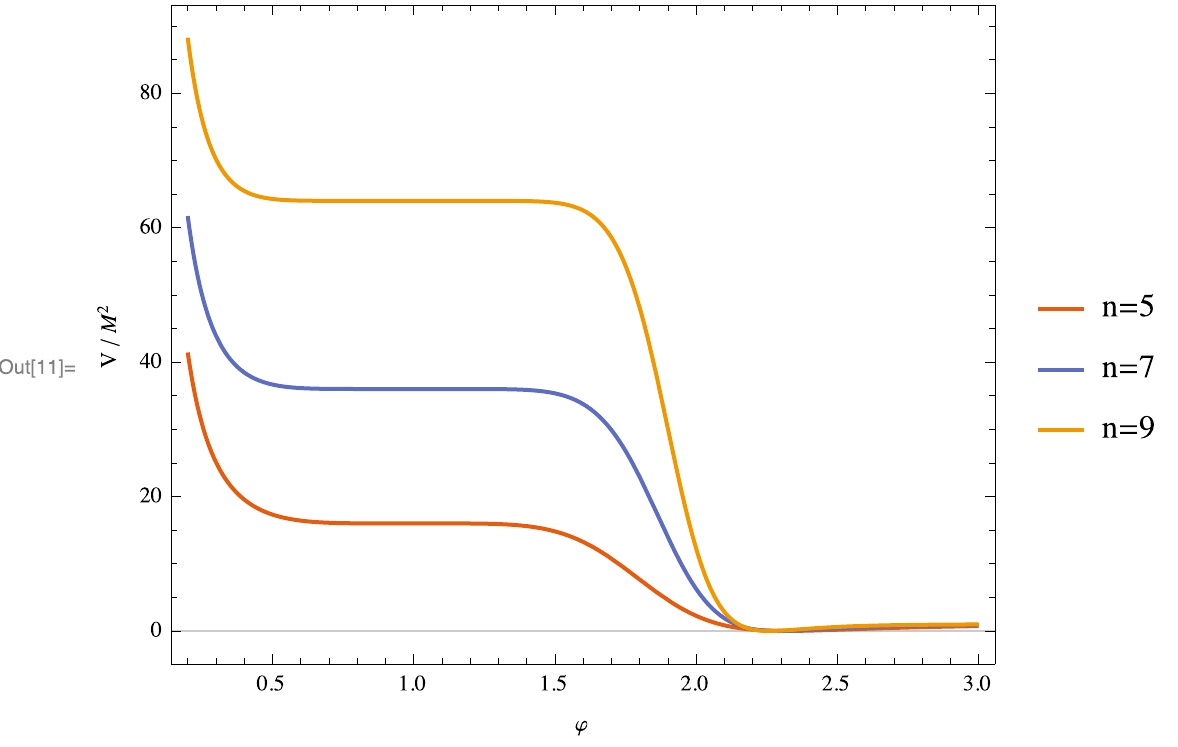}
\hspace{0.5cm}
\includegraphics[height=4.5cm]{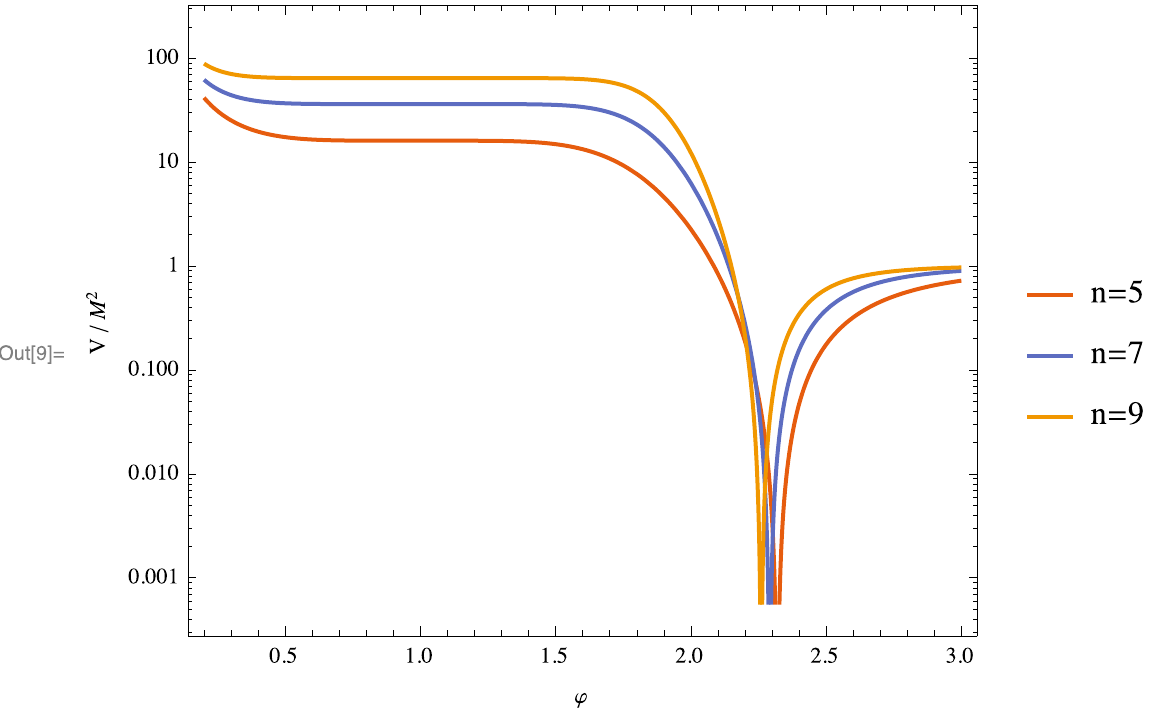}
\caption{\it Einstein Frame potential as a function of Jordan frame field for $\lambda = \lambda_1$, $\xi = 1$ and $\Lambda = 0$ in linear and logarithmic scales (left and right panels respectively). The scale ratio between the saddle-point and Starobinsky plateaus is $(n-1)^2$. Thus, for $n\sim M^{-1} \gg 1$ one finds $V(\varphi_s)\sim M_p^4$. This theory however suffers from big values of $\tilde{\lambda}_k$ coefficients.} 
\label{fig:Saddle1}
\end{figure}

For every $\lambda \neq \lambda_2$ one obtains two inflationary regions - one Starobinsky-like plateau in $f \gg 1$ regime and one around the saddle point at $\varphi_s$. For $\lambda > \lambda_2$ two plateaus are separated by the GR minimum, so the possible inflationary scenario is the following: 

\begin{itemize}

\item[i)] The first stage of inflation starts at the higher plateau, which is usually the one with the saddle-point. If $V(\varphi_s) \gg M^2$ then this stage may be consider as the pre-inflation, where the universe is homogenised at very high scales. In such a case scales of initial conditions for the universe ($M_p^4$) and of pre-inflation may be very close to each other, which solves (or at least weakens) the problem of initial conditions for inflation.

\item[ii)] After the pre-inflation ends the field roll down on the steep slope towards the GR minimum and overshoots it.

\item[iii)] The field rolls up on the second plateau and starts the second inflationary era. This time the scale of plateau is set by normalisation of primordial inhomogeneities. 

\item[iv)] Inflation finally ends when the field rolls towards the GR minimum. Potentials that could realize this kind of inflationary scenario are plotted in Fig. \ref{fig:Saddle1} (all values of $n$) and \ref{fig:Saddle2} (for $\lambda > \lambda_2$).
\end{itemize}
For $\lambda < \lambda_2$ the saddle-point plateau comes right after the Starobinsky one, without a minimum in between (see Fig. \ref{fig:Saddle2} for details). Thus, the evolution of $\varphi$ has again 4 stages:

\begin{itemize}

\item[i)] Inflation starts at the Starobinsky plateau, which scale may be much bigger than the GUT scale. As in the previous case, this may be used to solve the problem of fine tuning of initial conditions for inflation. 

\item[ii)] There is an inflationary break due to the fast-roll evolution on the steep slope between. plateaus.

\item[iii)] The second stage of inflation happens on the saddle-point plateau.

\item[iv)] Inflation finally ends when the field oscillates around the GR vacuum.
\end{itemize}

{The multi-phase inflation could be the source of primordial black holes due to the growth of primordial inhomogeneities in-between phases of inflation \cite{Clesse:2015wea}. Black holes could in principle play a role of barionic dark matter and therefore our model could be useful in explaining late-time evolution of the Universe. This issue shall be analysed in our further work.}

Even if the higher plateau is at the GUT scale the lower one can still generate some e-folds of inflation. Note, that due to the suppression of inhomogeneities at smaller scales we currently observe $\sim 15$ e-folds of the last $\sim 60$ e-folds of inflation. Therefore the last stage of inflation could happen in much lower scales, as long as it does not violate constrains on the CMB power spectrum.  
\\*

\begin{figure}[h]
\centering
\includegraphics[height=4.9cm]{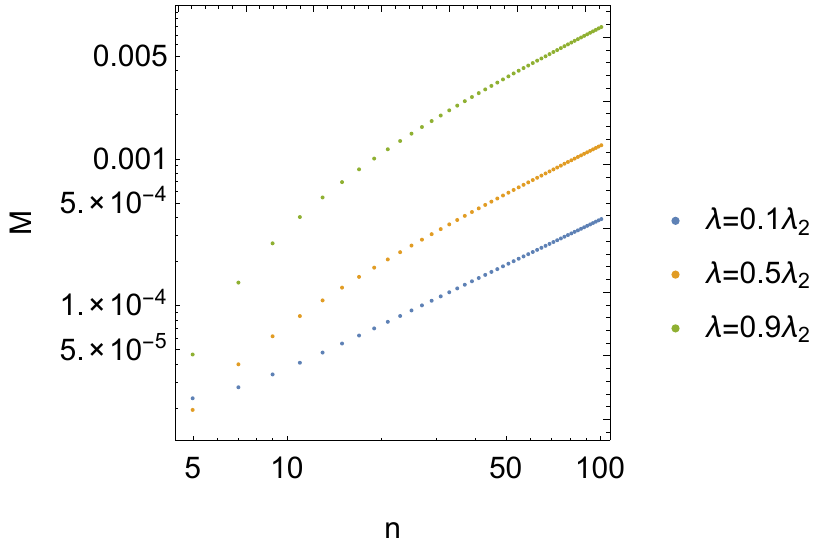}
\hspace{0.1cm}
\includegraphics[height=4.9cm]{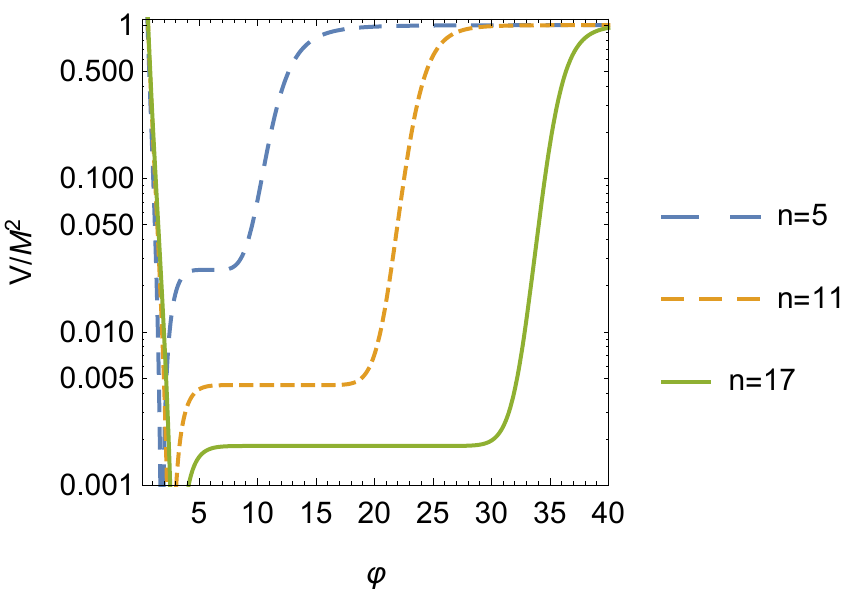} \\
\vspace{0.3cm}

\includegraphics[height=5cm]{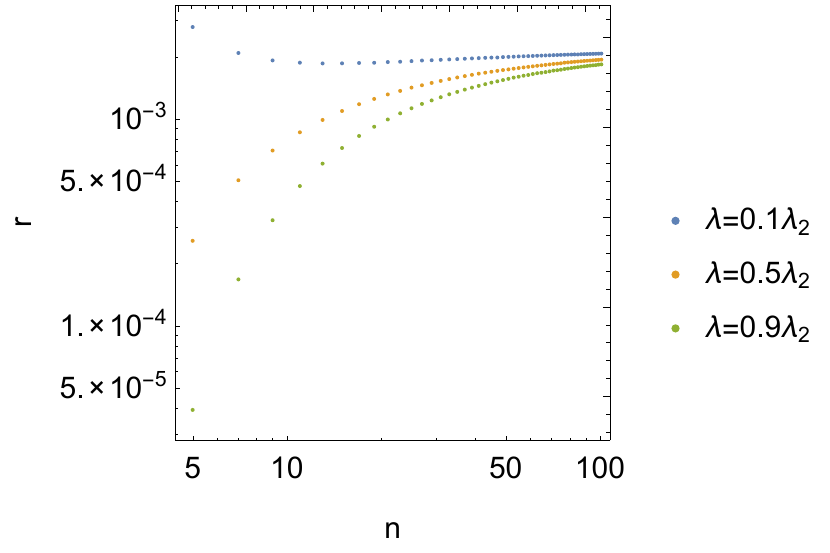}
\hspace{0.1cm}
\includegraphics[height=5cm]{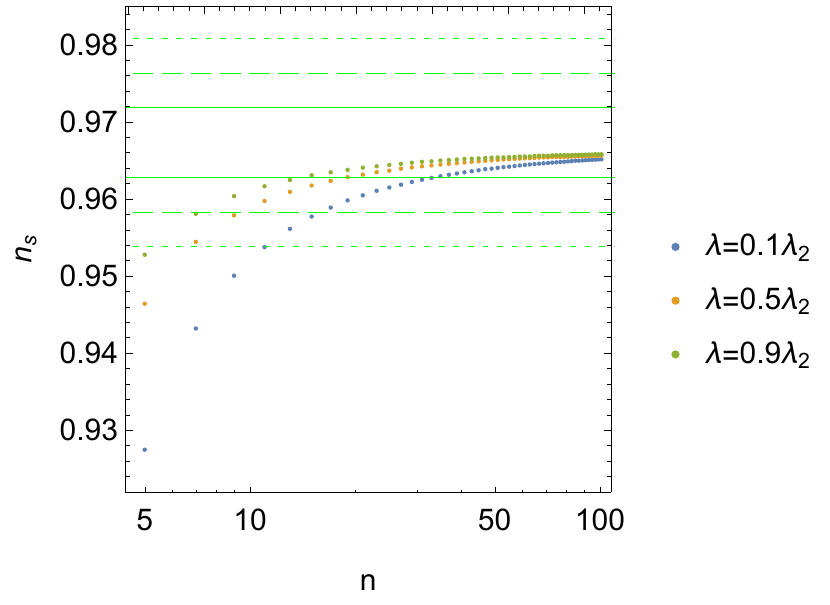}
\caption{\it Numerical results for $\Lambda = 0$ (Brans-Dicke-like), $\xi = 1$, and $\lambda \sim \lambda_1$. Solid, dashed and dotted green lines denote $1\sigma$, $2\sigma$ and $3\sigma$ areas of the Planck data respectively. The final stage of the inflation happens close to the saddle-point, which scale is much lower than the Starobinsky plateau. Note that for small $n$ the $r$ decreases rapidly for $\lambda \to \lambda_2$, which enables arbitrarily low scale of inflation for $|\varepsilon| \ll 1$.} 
\label{fig:rnsSaddle}
\end{figure}

The Brans-Dicke-like scenario includes the effect of lowering scale of the saddle point, which can be obtained e.g. by choosing a certain form of $\lambda$. From Eq. (\ref{eq:VsBD}) one finds, that for $\lambda = (1\pm\varepsilon)\lambda_2$ (where $\varepsilon$ is a small, positive constant) the scale of the saddle point is equal to
\begin{equation}
V(\varphi_s) = \left(\frac{M \, \varepsilon}{n-1}\right)^2 \label{eq:Vslambda2}
\end{equation}
and therefore can be arbitrarily low for sufficiently low value of $\epsilon$. This result is particularly interesting because $V(\varphi_s)$ does not depend explicitly on $\xi$. The Eq. (\ref{eq:Vslambda2}) suggests, that increasing the value of $n$ could also decrease the scale of the saddle-point inflation. This is not the case due to the strong $n$-dependence of $M$. Numerical results for $\lambda \sim \lambda_2$ are plotted in Fig. (\ref{fig:rnsSaddle}). The most important conclusion from those results is, that for sufficiently big $n$ all results fit to the Planck data. As predicted, taking small values of $\varepsilon$ decreases the scale of inflation, which in principle can be arbitrarily low. This fact, together with huge number of e-folds is particularly useful feature of the model - this kind of inflation is required for e.g. Higgs-Axion cosmological relaxation \cite{Espinosa:2015eda,Graham:2015cka}. 
\\*

Particularly interesting is the growth of $M$ with respect to $n$. The $M^2$ is the scale of the Starobinsky Plateau, so for sufficiently big $n$ the scale of the pre-inflation can reach the Planck scale (see upper-right panel of the Fig. \ref{fig:rnsSaddle} for details). This can be used to solve problem of initial conditions for inflation. Note that for given $\xi$ and $n \gg 1$ one obtains the same result on the $(n_s,r)$ plane independently from the value of $\lambda$ (see lower panels of the Fig. \ref{fig:rnsSaddle}). This attractor behaviour is significantly different than the one obtained in Ref. \cite{Kallosh:2013tua}, where the Authors obtain attractor behaviour in the $\xi \gg 1$ limit.

\section{Higgs inflation-like induced inflation} \label{sec:Higgs}

In order to investigate saddle-point generalisations of Higgs inflation-like induced models let us investigate the $\Lambda = 1$ scenario. Note that for $f(\varphi) = 1 + \xi \varphi^2$ one reconstructs the Higgs inflation. The main similarities and differences comparing to the Brans-Dicke-like case are following
\begin{itemize}

\item For any odd $n$ the GR vacuum of the theory is positioned in $\varphi = 0$. The result is independent of $\xi$ and $\lambda$

\item For $\varphi = \varphi_s$ one finds 
\begin{equation}
V_s =  M^2 \left(\frac{\xi}{\xi + n (n\, \lambda)^{\frac{1}{n-1}}}\right)^2
\end{equation}
and therefore one cannot obtain $V_s = 0$ for any $\lambda$ or $\xi$ besides the trivial case of $\xi = 0$. On contrary to the Brans-Dicke-like model the saddle point cannot have a higher scale than the Starobinsky one. Thus, the pre-inflation cannot appear in scales much higher than the GUT scale and the $\Lambda = 1$ scenario cannot be used in the context of weakening or solving the problem of initial conditions.

\end{itemize}

\begin{figure}[h]
\centering
\includegraphics[height=5cm]{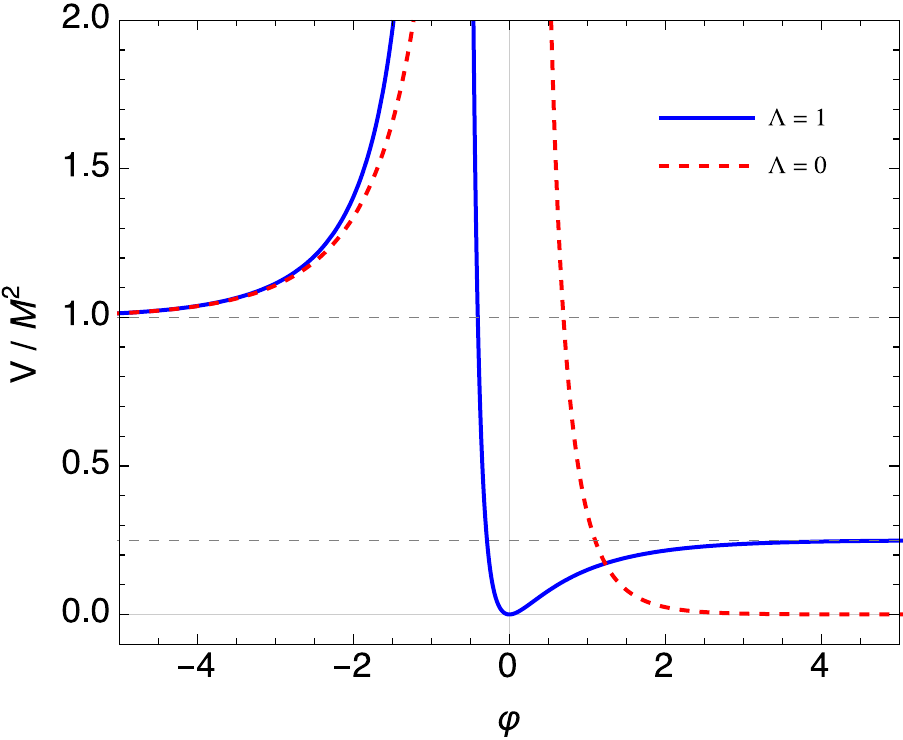}
\hspace{0.1cm}
\includegraphics[height=5cm]{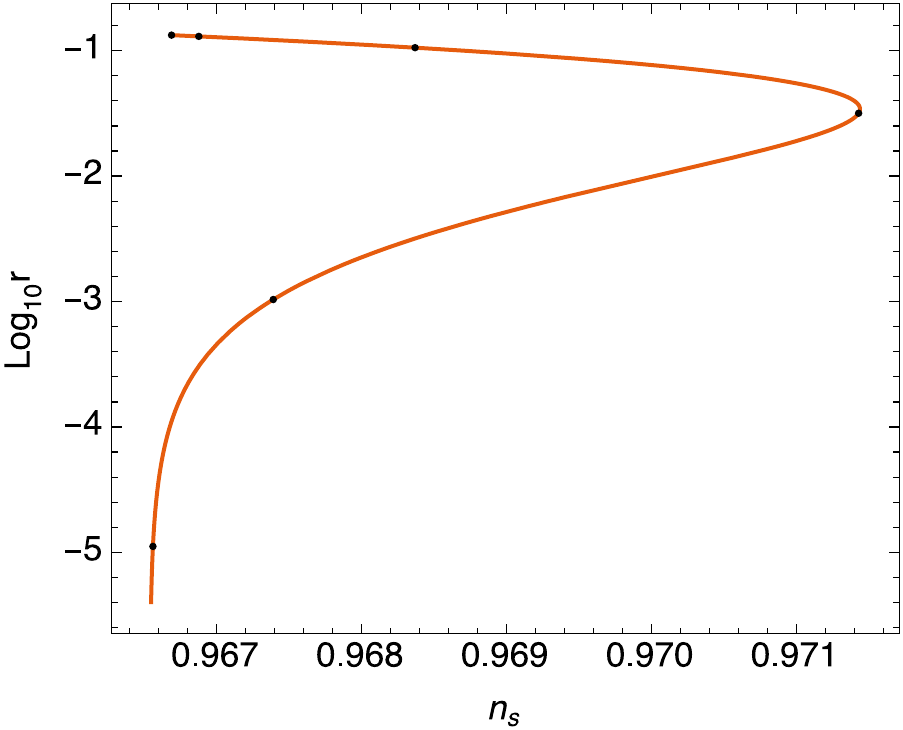}
\caption{\it Left panel: Einstein frame potentials for $\lambda \propto \lambda_2$, $n \to \infty$, and for $\Lambda = 1$ or $\lambda = 0$. Note that only in the first case gives the inflationary plateau and a GR vacuum. Right panel: The $(n_s,r)$ plane for the (\ref{eq:ninf}) model with $\Lambda = 1$. Dotes denote $\xi=\{10^{-4},10^{-3},10^{-2},10^{-1},1,10\}$ (from the top to the bottom). All results are within the $2\sigma$ of the PLANCK data. The maximal considered is $\xi \sim 17$.} 
\label{fig:rnsSaddleHninf}
\end{figure}

Numerical results for the saddle-point inflation for $\lambda \propto \lambda_2$, $\Lambda = 1$ and $n \to \infty$ are plotted in Fig. \ref{fig:rnsSaddleHninf}. The results are consistent with the Planck data and in the strong coupling regime one obtains rather small scale of inflation. The maximal value of $\xi$ is of order of $17$, which is much smaller than in the case of regular Higgs inflation, for which $\xi \lesssim 10^4$.

The (\ref{eq:ninf}) model gives the correct inflationary potential even in the case of the minimal coupling to gravity, i.e. when the gravitational part of the action takes the form of $\frac{1}{2}\int d^4 x \sqrt{-g} R $. For $\Lambda = 1$ one finds 
\begin{equation}
U(\varphi) = M^2\left(1-e^{-\xi \varphi}\right)^2 \, ,
\end{equation}
which is the Einstein frame potential for the Brans-Dickie generalisation of the Starobinsky inflation, where 
\begin{equation}
\omega_{\text{\tiny BD}} = \frac{1}{\xi^2} - \frac{3}{2} \, .
\end{equation}
This result strongly correspond to the result from \cite{Artymowski:2015pna,Artymowski:2016mlh}, where we have considered $f(R) = \sum_{k=1}^n \alpha_k R^k$ model. In the $n \to \infty$ limit the requirement of the maximal flatness gave the Starobinsky model with exponentially suppressed correction. {Nevertheless it is important to note that the scalar approach gave the 
different Einstein frame potential than requiring extreme flatness in the pure $f(R)$. In the $f(R)$ approach one cannot even write the Ricci scalar as an analytical function of the Einstein frame field and therefore the Einstein frame potential can be expressed only as a function of $R$.}


\section{Summary} \label{sec:Summary}

In this paper we present the model of the induced inflation with two plateaus and therefore with two scales of inflation. In Sec. \ref{sec:general} we require the existence of a stationary point of the Einstein frame potential $V$ for any function {describing} a non-minimal coupling to gravity denoted as $f(\varphi)$. We apply this requirement to $f = \xi \sum_{k=0}^{n} \lambda_k \varphi^k$ and we obtain the simplified form of $f(\varphi)$ with 4 free parameters: $\Lambda := \lambda_0$, $\lambda := \lambda_n$, $\xi$ and $n$. We show that besides of the Starobinsky plateau, $V(\varphi)$ has a saddle point, which generate inflation with different energy scale. The saddle point depends only on $n$ and $\lambda$. We note that $f$ takes simplified forms for two particular forms of $\lambda$, namely for $\lambda_1: = 1/n$ and $\lambda_2 := (\xi/n)^n/\xi$. In the letter case there exists the $n \to \infty$ limit, which gives $f(\varphi) = 1+ \Lambda - e^{-\xi \varphi}$. The requirement of the perturbativity of the Jordan frame potential favours $\lambda \sim \lambda_2$ over $\lambda \sim \lambda_1$. 

{In Sec. \ref{sec:alphaattractors} we discuss the result of Ref. \cite{Artymowski:2016pjz}, namely that the flatness of the potential, which appears due to the existence of the stationary point at $\phi_s$ can be also expressed in terms of $\alpha$-attractors. If one defines $f$ as an inflaton, then its kinetic term obtains a pole at $f(\varphi_s)$. The kinetic term is similar to the one from the $\alpha$-attractors and {in the $n \to \infty$ limit} it can take exactly the same form after another redefinition of field.}
\\*

In Sec. \ref{sec:BD} we perform numerical and analytical analysis of the Brans-Dicke-like induced inflation, i.e. with $\Lambda = 0$. For $\lambda = \lambda_2$  one obtains only one plateau (the Starobinsky one) and the  stationary point is the GR minimum, around which the potential is very flat, but not inflationary. In general two plateaus can be separated by the GR vacuum or a cascade. In both cases GR vacuum is the only vacuum of the model. For $\lambda \sim \lambda_2$ the saddle-point plateau has a very low scale and for sufficiently big $n$ such a saddle point inflation is fully consistent with the PLANCK data. This kind of inflation can be arbitrarily long lasting and low scale, so it can be used in the context of Higgs-Axion relaxation. The other way to obtain big difference between scales of inflation is to consider the $n \gg 1$ limit, for which the Starobisnky plateau can have arbitrarily high scale, e.g. the Planck scale. Pre-inflation could homogenise the Universe at the Planck scale and solve the problem of initial conditions for inflation. For $\lambda = 0$, $\lambda = \lambda_2$ and $n \to \infty$ the Einstein frame potential does not have a GR minimum in the attractive gravity regime, so this case becomes unphysical. 
\\*

In Sec. \ref{sec:Higgs} we analyse the Higgs inflation-like scenario, for which $\Lambda = 1$. The main difference comparing to the Brans-Dicke-like model is that besides the trivial case of $\xi = 0$ one always obtain two plateaus, even for $\xi = n$. The Starobinsky plateau is always higher than the scale of the saddle point, which means that the pre-inflation cannot solve problem of initial conditions for inflation unless $M \sim \mathcal{O}(1)$. For $\lambda \propto \lambda_2$ the $b \to \infty$ limit gives an inflationary model with perfect consistency with the Planck data. The GR vacuum of the Einstein frame potential is separated from the repulsive gravity regime by the infinite wall of the potential. Taking the $\Lambda = 1, \lambda \propto \lambda_2, n\to \infty$ model for the minimal coupling to gravity gives the Einstein frame potential of the Brans-Dicke generalisation of the Starobinsky inflation.

\acknowledgments

MA would like to thank Javier Rubio for useful discussions. This work was partially supported by the National Science Centre under research grants DEC-2012/04/A/ST2/00099 and DEC-2014/13/N/ST2/02712.
ML was supported by the Polish National Science Centre under doctoral scholarship number 2015/16/T/ST2/00527. MA was supported by National Science Centre grant FUGA UMO-2014/12/S/ST2/00243. 
ZL thanks DESY Theory Group for hospitality. This work was supported by the German Science Foundation (DFG) within the Collaborative Research Center (SFB) 676 Particles, Strings and the Early Universe.\\

\end{document}